\documentclass[a4paper,11pt]{article}
\pdfoutput=1 

\usepackage{jheppub} 
\usepackage[T1]{fontenc} 
\usepackage[dvipsnames]{xcolor}
\usepackage{amsmath}
\usepackage{amssymb}
\usepackage{amsfonts}
\usepackage{comment}
\usepackage{enumerate}
\usepackage{changes}
\usepackage{color}
\usepackage{multirow}
\usepackage{graphicx}

\definecolor{darkgray}{gray}{0.25}
\definecolor{darkgreen}{rgb}{0,0.5,0}

\newcommand{\beq}{\begin{equation}}
\newcommand{\eeq}{\end{equation}}

\def\be{\begin{equation}}
\def\ee{\end{equation}}
\def\bea{\begin{eqnarray}}
\def\eea{\end{eqnarray}}

\title{A collider observable QCD axion}

\author{Savas Dimopoulos,}
\author{Anson Hook,}
\author{Junwu Huang,}
\author{Gustavo Marques-Tavares}

\affiliation{Stanford Institute for Theoretical Physics, Stanford University, Stanford, CA 94305, USA}

\emailAdd{savas@stanford.edu}
\emailAdd{hook@stanford.com}
\emailAdd{curlyh@stanford.edu}
\emailAdd{gusmt@stanford.edu}

\abstract{
We present a model where the QCD axion is at the TeV scale and visible at a collider via its decays.  Conformal dynamics and strong CP considerations account for the axion coupling strongly enough to the standard model to be produced as well as the coincidence between the weak scale and the axion mass.   The model predicts additional pseudoscalar color octets whose properties are completely determined by the axion properties rendering the theory testable.}

\begin{document} 
\maketitle

\flushbottom

\section{Introduction}

The QCD axion~\cite{Peccei:1977hh,Peccei:1977ur,Weinberg:1977ma,Wilczek:1977pj} is arguably the simplest and most elegant solution to the strong CP problem.  The axion as originally proposed had a mass in the keV range and was excluded soon afterwards due to astrophysical and cosmological constraints~(see e.g. Ref.~\cite{Essig:2013lka} and references therein).  The two approaches to evading these constraints have been to make it heavier~\cite{Dimopoulos:1979qi, Dimopoulos:1979pp} or to make it more weakly coupled~\cite{Kim:1979if,Shifman:1979if,Zhitnitsky:1980tq,Dine:1981rt}.  In this work, we take the approach of Ref.~\cite{Dimopoulos:1979qi, Dimopoulos:1979pp} and show that the QCD axion can be heavy and strongly coupled enough for it to be directly observed at ATLAS and CMS.

In the interest of minimality we work in a framework where the cosmological constant and hierarchy problems are addressed by anthropic selection. This is the only known approach to the cosmological constant problem. It eliminates the need for either supersymmetry or compositeness in the electroweak sector and concentrates on the strong CP problem for which there is no anthropic solution.

The phenomenology of our model is extremely predictive.  The lightest particles are the axion and a slightly heavier scalar color octet.  The mass and couplings of the adjoint state are predicted from the axion properties up to a discrete choice of hypercharge quantum numbers.  Another distinct feature of our model is the absence of additional pseudo-goldstones charged under the electroweak gauge groups with masses near the axion, which are ubiquitous in models with a single confining sector (see e.g. Ref~\cite{Harigaya:2015ezk}).

\section{Model}

In order for the QCD axion to be collider visible, there are several constraints that any model must satisfy.  Firstly, we must increase the mass of the axion in such a way that does not reintroduce the strong CP problem.  Secondly, we must explain why the axion is so strongly coupled, i.e. $f_a \sim m_a$, which is needed so that it can be produced with an observable cross section at a collider.  Thirdly, we must do all of this without introducing any new hierarchy problems.  Finally, we need to motivate why all of these dynamics happen around the TeV scale.

Increasing the mass of the axion requires another gauge group with the same theta angle as QCD.  As was pointed out in Ref.~\cite{Hook:2014cda}, $\mathbb{Z}_2$ solutions to the strong CP problem motivate new particles and a composite axion around the TeV scale.
We are thus led to consider a $\mathbb{Z}_2$ copy\footnote{When referring to copies it is more natural to use the permutation group, but in an effort to use the same terminology as the rest of the community, we will refer to our symmetry as a $\mathbb{Z}_2$ symmetry.} of the Standard model with a shared axion~\cite{Rubakov:1997vp,Berezhiani:2000gh}.
In order for the axion mass to be significantly different from the usual QCD axion, the confinement scale of the mirror QCD must be significantly larger than our QCD scale.  This is accomplished by softly breaking the $\mathbb{Z}_2$ symmetry by giving their Higgs a much larger vacuum expectation value, e.g. $10^{14}$ GeV.
Because $f_a$ must be at the TeV scale in order to be visible at a collider, we consider a composite axion in order to avoid any additional hierarchy problem.  It turns out that a simple model satisfying these two criteria also naturally predicts that $f_a \sim m_a$, which, as we will discuss, cannot be much beyond the TeV scale.

\begin{figure}[t]
\begin{center}
\includegraphics[width=0.4\textwidth]{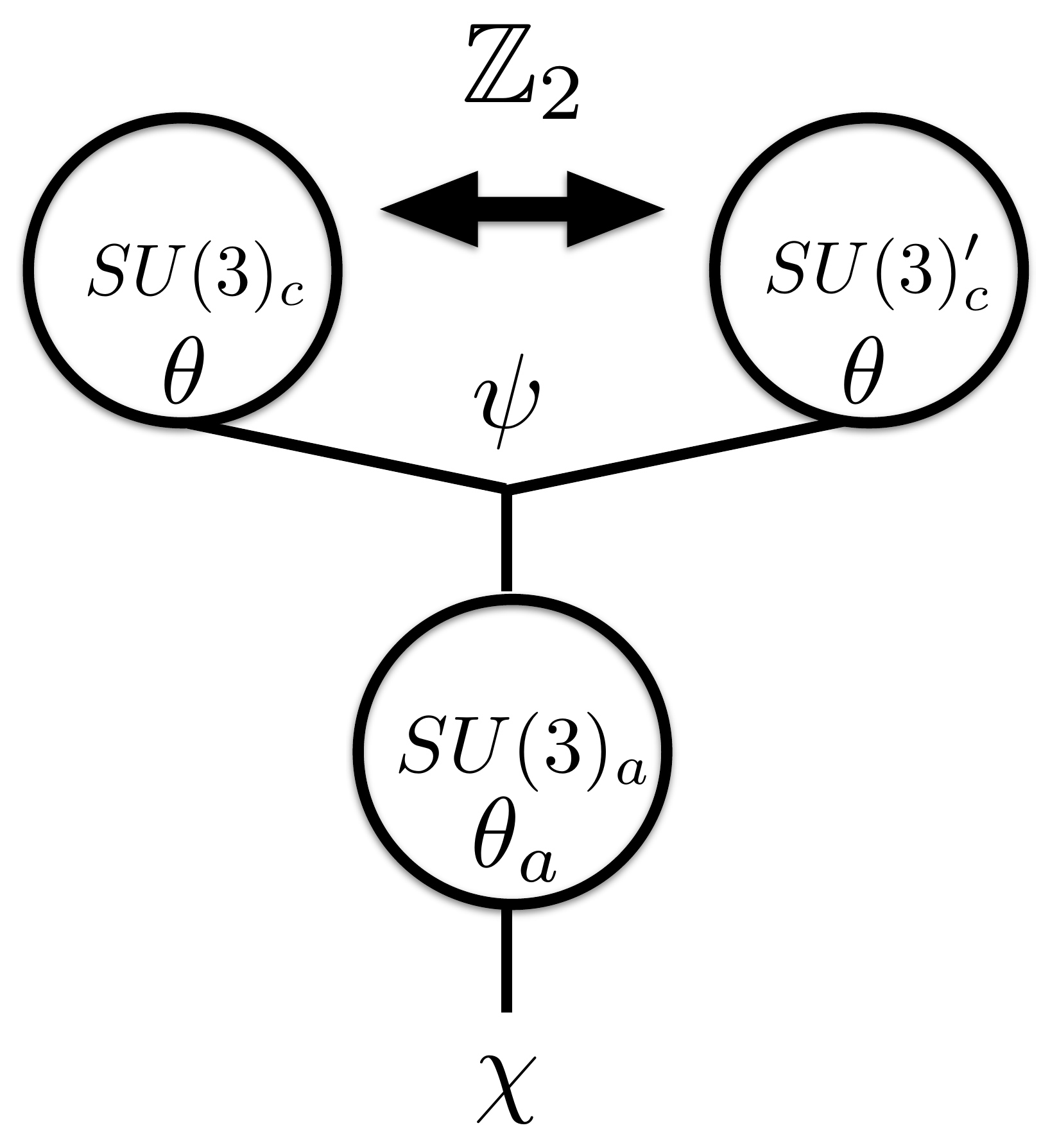}
\caption{A pictorial representation of our model.  There are two copies of the Standard model related by a $\mathbb{Z}_2$ symmetry with a shared composite axion.  All theta angles can be set to zero in the UV by using chiral rotations of $\psi$ ($\psi^c$) and $\chi$ ($\chi^c$).} \label{Fig: model}
\end{center}
\end{figure}

Our model is summarized in Fig.~\ref{Fig: model}.   There are two copies of the Standard model related by a $\mathbb{Z}_2$ symmetry with a shared U(1) gauge boson.  This symmetry is softly broken by differing Higgs masses\footnote{Alternatively, if the Higgs potential has multiple minima, the symmetry could be spontaneously broken by the Higgses living in different vacua~\cite{Blinov:2016kte}.}.  The mirror Higgs has a large vev, which breaks $SU(2)' \times U(1)$ down to $U(1)_Y$ and gives all mirror fermions large masses. Therefore all mirror particles are integrated out at a large scale leaving only the gluons of QCD$'$ at low scales.  Because the mirror quarks have been integrated out at high energies, QCD$'$ confines at a scale higher than QCD.  After the mirror quarks have been integrated out, QCD and QCD$'$ in principle can have different $\theta$ angles but RG effects in these types of models are small~\cite{Ellis:1978hq}.

A composite axion is added to the theory by adding a new gauge group $SU(3)_a$ with vector-like tri-fundamentals $\psi$ and $\psi^c$ under $SU(3)_a \times SU(3)_c \times SU(3)'_c$ as well as a vector-like pair of $SU(3)_a$ fundamentals $\chi$ and $\chi^c$ as illustrated in Figure~\ref{Fig: model}.  When $SU(3)_a$ confines, it leads to the symmetry breaking pattern $U(10)_L \times U(10)_R / U(10)_D$.  This results in a 100 pseudo-goldstones which are charged under $SU(3)_c \times SU(3)'_c$ as 
\bea
\pi_A = (8 , 1) \qquad & \pi'_A = (1 , 8) \qquad  \Pi_A = (8 , 8)  \qquad \eta'_A = (1 , 1) \nonumber \\
 \phi_A = (3, \overline 3)  &  \overline{\phi}_A = (\overline 3, 3)  \qquad a = (1 , 1)
\eea
with all particles other than $\eta'_A$ and $a$ obtaining a mass from gauge boson loops.
At 1 loop the charged goldstones get masses
\beq
\begin{aligned}
	m^2_{\pi_A} \approx \frac{9 \alpha_s}{4\pi} \Lambda_a^2, & \qquad m^2_{\pi'_A} \approx \frac{9 \alpha'_s}{4\pi} \Lambda_a^2 , \\
	m^2_{\phi_A} =m^2_{\overline{\phi}_A} \approx \frac{3 (\alpha'_s + \alpha_s)}{8\pi} \Lambda_a^2  , & \qquad m^2_{\Pi_A} \approx \frac{9 (\alpha'_s + \alpha_s)}{4\pi} \Lambda_a^2,
\end{aligned}
\eeq
where $\Lambda_a$ is the confinement scale of the $SU(3)_a$ group. The $\eta'_A$ obtains a mass from the $SU(3)_a$ anomaly while $a$ is our composite axion.  
Due to the tri-fundamental nature of $\psi$, $a$ is a shared axion between QCD and QCD$'$.  If $SU(3)_a$ confines at a high scale, then the only IR degree of freedom is the composite axion $a$.  However, if the QCD axion is to be visible at low energies, $f_a$ must be near the TeV scale and thus some of the other pseudo-goldstone bosons (the lightest being $\pi_A$) might be accessible at a collider.

More explicitly, we work with the following representation for the non-linear sigma model associated with the heavy axion:
\beq
\Sigma_a = e^{ 2 i a T_{\rm PQ} / f_a},
\eeq
where $T_{\rm PQ}$ is the generator associated with the Peccei-Quinn symmetry
\beq
T_{\rm PQ} = \frac{1}{\sqrt{180}} \left(
	\begin{array}{cc}
	\mathbf{1}_{9\times9} & 0 \\
	0 & -9
	\end{array}
\right) .
\eeq

Since $\psi$ and $\chi$ are $SU(2)$ singlets, the coupling of the axion to the QCD and QCD$'$ anomaly, with the above convention, are given by
\beq
\label{eq: ggdual-coupling}
\mathcal{L} \supset  \frac{3 a}{64\sqrt{5} \pi^2 f_a} \epsilon_{\mu \nu \rho \sigma} \left( g_3^2 G^{a \mu \nu} G^{a \rho \sigma} + g_3^{\prime 2} G^{\prime a \mu \nu} G^{\prime a \rho \sigma} \right) .
\eeq 
The coupling to the Hypercharge gauge bosons is
\beq
\label{eq: bbdual-coupling}
\mathcal{L} \supset - \frac{9}{\sqrt{5}} \frac{g_1^2 a}{32 \pi^2 f_a} \left(Y_\psi^2 -Y_\chi^2 \right)  \epsilon_{\mu \nu \rho \sigma} B^{\mu \nu} B^{\rho \sigma} .
\eeq
The couplings of the light adjoint scalar $\pi_A$ are
\beq
\label{eq: octetdual-coupling}
\mathcal{L} \supset \frac{1}{2}(D_{\mu} \pi_A^a)(D_{\mu} \pi_A^a) +  \frac{3 \sqrt{3} g_3^2 \pi_A^a}{64 \pi^2 f_a} d_{abc}  \epsilon_{\mu \nu \rho \sigma} G^{b \mu \nu} G^{c \rho \sigma} + \frac{3 \sqrt{3} g_3 g_1 Y_{\psi} \pi_A^a}{16 \pi^2 f_a}  \epsilon_{\mu \nu \rho \sigma} G^{a \mu \nu} B^{ \rho \sigma} ,
\eeq
where $d^{abc} = 2 \text{Tr}(T^a \{T^b , T^c\})$. 

After confinement of QCD$'$, the coupling in Eq.~\ref{eq: ggdual-coupling} gives the axion a mass.  
Using the results of Witten~\cite{Witten:1979vv}, Veneziano~\cite{Veneziano:1979ec}, the large N limit, the observed results for QCD and rescaling the energy densities involved by $\Lambda_{\rm QCD'}/\Lambda_{\rm QCD}$, we can calculate the contribution to the axion mass from $\Lambda_{QCD'}$ to be
\bea
\label{eq: axion mass}
m^2_a = \frac{9}{5 f_a^2} \left(\frac{\Lambda_{\rm QCD'}}{\Lambda_{\rm QCD}}\right)^4 \frac{ (m^2_{\eta'}+m^2_\eta - 2 m_K^2) f_\pi^2}{6} .
\eea
If $\Lambda_{QCD'} < \Lambda_{QCD}$, then the standard QCD contribution to the axion mass is more important.  The rescaling $\Lambda_{\rm QCD'}/\Lambda_{\rm QCD}$ has some amount of uncertainty due to $\Lambda_{\rm QCD}(\Lambda_{\rm QCD'})$ being very sensitive to the number of loops used in its determination.

At this point, an observant reader might realize that when treating QCD and QCD$'$ as flavor groups, $SU(3)_a$ might not actually be a confining theory.  In particular, for a single $SU(3)$ gauge group, the conformal/confining transition happens somewhere between $8 < N_f < 12$~(see Ref.~\cite{Neil:2012cb} and references therein).  We will take as an {\it assumption} that $N_f = 10$ is conformal while $N_f = 9$ is confining. In this case, if $SU(3)_a$ becomes strongly coupled long before QCD$'$, it reaches an approximate fixed point without chiral symmetry breaking.  If $N_f = 10$ is not conformal and is instead confining, then to have a sizable collider cross section we require a coincidence of scales between the confinement of $SU(3)_a$ and $SU(3)_{\rm QCD'}$.

To examine what happens as a function of RG scale, we first set $\alpha_s = 0$ and study what happens with QCD$'$ and $SU(3)_a$.  When $\alpha'_s = 0$, $SU(3)_a$ does not confine and instead approaches a conformal fixed point.  When $\alpha'_s = \infty$, the tri-fundamental $\psi$ have been removed from the IR by the QCD$'$ confinement leaving only one flavor charged under $SU(3)_a$ which now confines. Thus, we find that if $SU(3)_a$ sits at its CFT fixed point in the UV, it will confine and break chiral symmetry only when QCD$'$ becomes strong!  We have found a dynamical reason why $SU(3)_a$ and QCD$'$ confine around the same scale!  This approach is similar to attempts to link the QCD and EW scales in technicolor~\cite{Appelquist:1997fp} as well as tumbling gauge theories~\cite{Raby:1979my}.

\begin{figure}[t]
\begin{center}
\includegraphics[width=0.5\textwidth]{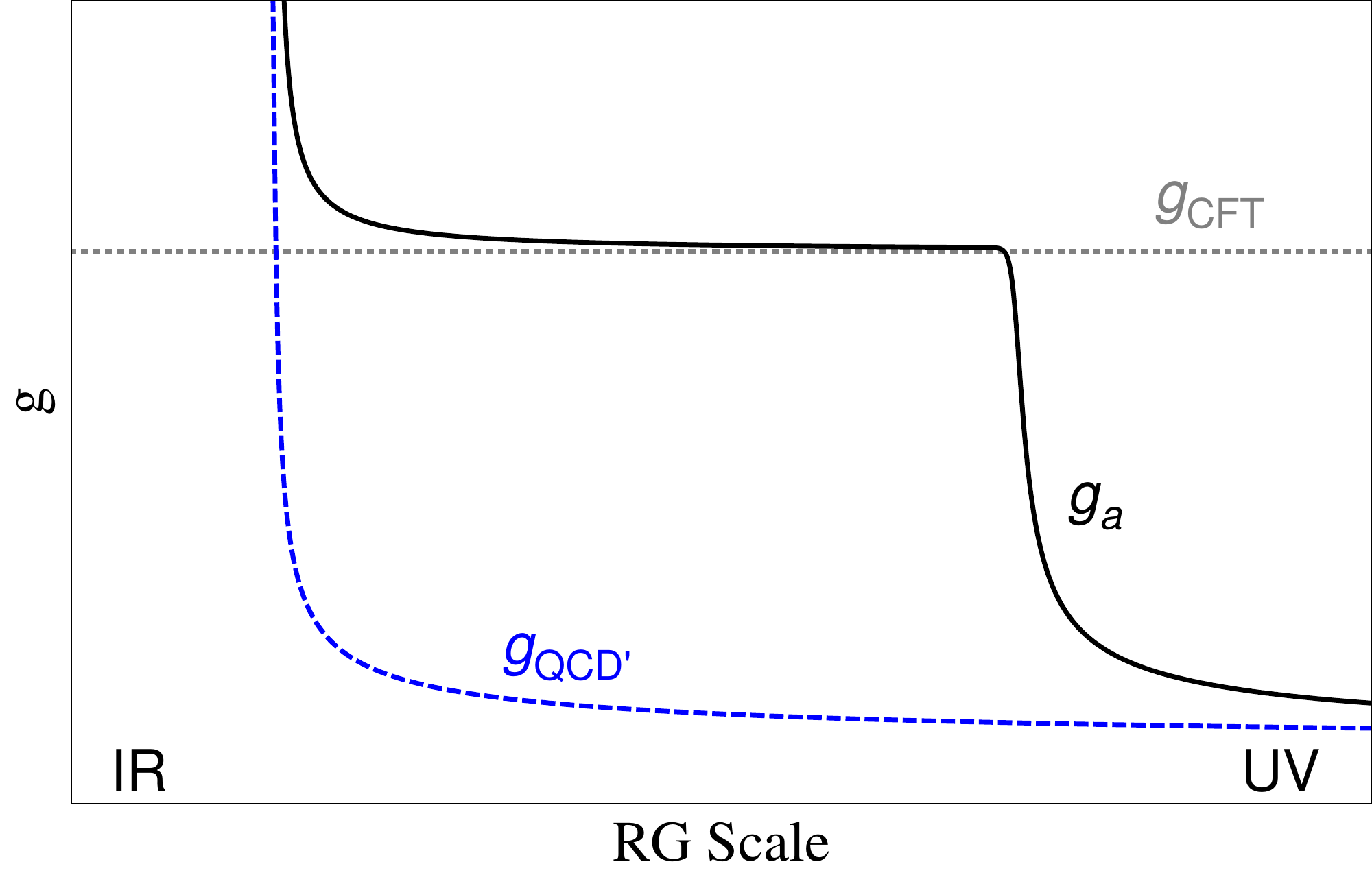}
\caption{A pictorial representation of how we imagine the confinement of QCD$'$ (Blue, Dashed) triggers the confinement of $SU(3)_a$ (Black, Solid).  We know for certain that in the IR both gauge theories have confined and in the UV that $SU(3)_a$ has reached a conformal fixed point, but in between is a picture of how we imagine things will proceed.  Lattice calculations are needed to find which gauge theory confines first or what is the ratio of their confinement scales.} \label{Fig: gauge}
\end{center}
\end{figure}

While this approach is incalculable without lattice help, we show how this approach might plausibly work in Fig.~\ref{Fig: gauge}.  We know that when QCD$'$ is weakly coupled that $SU(3)_a$ flows to a CFT fixed point. Then as QCD$'$ becomes strong, the QCD$'$ and $SU(3)_a$ gauge couplings can lead to sizable contributions to each others beta functions. Then, according to the previous discussion, the two gauge groups should confine at similar scales, as depicted in Fig.~\ref{Fig: gauge}.

Finally, we discuss why QCD$'$ is expected to confine around the TeV scale or smaller.  Solutions to the strong CP problem are all constrained by higher dimensional operators. One $\mathbb{Z}_2$ symmetric operator that can cause problems is
\bea
 i \frac{H^\dagger H Y_u Q H u + H'^\dagger H'  Y_u Q' H' u'}{M_{\rm pl}^2} + h.c.,
\eea
where $M_{\rm pl}$ is the Planck scale. After a chiral rotation, this changes $\theta$ ($\theta'$) by $ H^\dagger H/M_{\rm pl}^2$ ($H'^\dagger H' /M_{\rm pl}^2$).  After the $\mathbb{Z}_2$ symmetry is broken by the two Higgses getting different vevs, this causes the two theta angles to be unequal.  Requiring that the difference be smaller than $10^{-10}$ gives us the constraint that
\bea
\langle H' \rangle \lesssim 10^{14} \, \text{GeV} .
\eea

Because the QCD and QCD$'$ gauge couplings were equal at high energies and RG evolve differently just because of when quarks are integrated out, we can translate the bound on $H'$ into a bound on the QCD$'$ confinement scale.  At one loop and taking $\Lambda_{a} \approx \Lambda_{\rm QCD'}$, the ratio of QCD to QCD$'$ dynamical scales depends on only $\langle H' \rangle$, $\langle H \rangle$ and yukawa couplings.  The incalculable piece coming from the strongly coupled particles $\psi$ cancel out in the ratio.   Thus at one loop we find that
\bea
\Lambda_{\rm QCD'} \sim \left ( \left( \frac{m_{t'} m_{b'} m_{c'} m_{s'} m_{d'} m_{u'} }{m_{t} m_{b} m_{c} } \right)^{2/3} \Lambda_{\rm QCD}^9 \right )^{1/11} \lesssim 10 \, \text{TeV},
\eea
which implies that the axion mass $m_a \lesssim 3\, \text{TeV}$ given Eq.~\ref{eq: axion mass}. This bound is really more of a guideline than an exact number as it relies on higher dimensional operators with incalculable coefficients and relies on a one loop estimate of the confinement scale.  At two loops, input from lattice is necessary to determine the effects of the CFT on the RG running of the QCD$'$ coupling constant.  Either way, this gives strong CP motivations for why the scale of new physics should be at the TeV scale. Even if we ignore higher-dimensional operators and take the $H'$ vev to the Planck scale, the confinement scale is less than $\sim 100 \,{\rm TeV}$.  The axion then has a mass less than $100 \, {\rm TeV}$ and would be visible at future colliders.

\section{Phenomenology}

\begin{figure}[t]
\begin{center}
\includegraphics[width=0.8\textwidth]{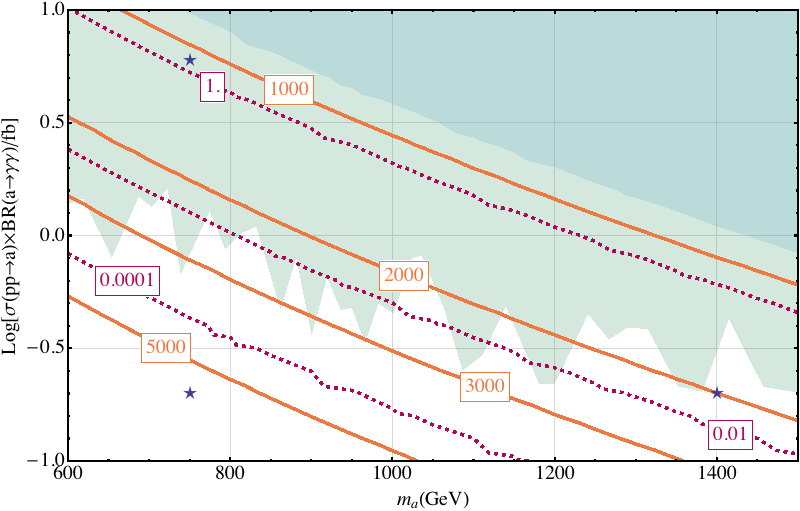}
\caption{We show a contour plot as a function of axion mass and production cross section times branching ratio into photons for the benchmark choice of hypercharge $(Y_{\psi},Y_{\chi}) = (0,1)$. The magenta dashed lines are the contours for the color octet $\pi_A$ production cross section at 13 TeV in pb ($1$, $10^{-2}$, $10^{-4}$) while the orange solid lines are the mass of $\pi_A$ in GeV (1000, 2000, 3000, 5000).  These two track each other fairly well because the cross section scales with the mass to a high power.  The blue shaded region is excluded by dijet searches for single $\pi_A$ production at 8 TeV~\cite{Aad:2014aqa,Khachatryan:2015sja}. The green region is excluded by direct diphoton searches for the axion decaying into pairs of photons~\cite{ATLAS:new,CMS:new}. The stars correspond to the benchmark points described in Table~\ref{table:masses of states}} \label{fig:axion-pion}
\end{center}
\end{figure}
In this section we describe the collider phenomenology of the model. The properties of the model are completely determined by the four parameters: $f_a, \, \Lambda_{\rm QCD'}, \, Y_\psi$ and $Y_\chi$. Although $f_a$ and $\Lambda_{\rm QCD'}$ are not truly independent, in practice we will treat them independently given the difficult in obtaining the precise relation between the two without a dedicated lattice investigation. We also use Eq.~\ref{eq: axion mass} to treat the axion mass $m_a$ as an input parameter instead of $\Lambda_{\rm QCD'}$.

The axion partial decay widths to gluons, photons and weak gauge bosons can be easily calculated at leading order using Eqs.~\ref{eq: ggdual-coupling} and \ref{eq: bbdual-coupling},
\beq
\begin{aligned}
	\label{eq: axion-partial-width}
	\Gamma(a \rightarrow gg )  & = \frac{9 \alpha_s^2}{160 \pi^3} \frac{m_a^3}{f_a^2} \, , \\
	\Gamma(a \rightarrow \gamma \gamma) = \frac{81 \alpha^2}{320 \pi^3} \frac{\left( Y_\psi^2 - Y_\chi^2 \right)^2 m_a^3}{f_a^2} & = \frac{\Gamma(a \rightarrow \gamma Z)}{2 \tan^2 \theta_w} =  \frac{\Gamma(a \rightarrow Z Z)}{\tan^4\theta_w} .
\end{aligned}
\eeq
From this we find that the branching ratio into the photons is a few percent for $Y\sim \mathcal{O}(1)$. The leading order production cross-section from gluon fusion is given by
\beq
\sigma(pp\rightarrow a) = \frac{\pi^2}{8 m_a s} \Gamma (a \rightarrow gg) \int_0^1 dx_1 \int_0^1 dx_2 \delta (x_1 x_2  - m_a^2/s) f_g(x_1) f_g(x_2),
\label{eq: axion-production}
\eeq
where $f_g$ is the gluon parton distribution function (PDF) and $s$ is the collider's energy squared. For a fixed choice of hypercharge assignments, the axion mass and its production times branching ratio to diphotons uniquely fix the other two parameters of the model $f_a$ and $\Lambda_{\rm QCD'}$, and subsequently the spectrum and collider signatures of the model.

Assuming that both $\psi$ and $\chi$ have hypercharges less than $\sim 2$, the axion decays predominantly into pairs of gluons. Due to the large dijet background, for most choices of hypercharge, the searches for diphoton, $ZZ$ and $Z\gamma$ final states are more sensitive to axion production than dijet searches. In fact, we see from Eq.~\ref{eq: axion-partial-width} that the decay to photons is significantly larger than to $Z\gamma$ and $ZZ$, which makes resonant diphoton searches the most promising probe of a collider visible axion. 
 
The scalar color octet $\pi_A$ is the second lightest state in our spectrum. $\pi_A$ has a mass
\beq
\label{eq: sigma-mass}
m_{\pi_A} \approx 2.3 f_a .
\eeq
We obtain this mass by simply scaling up the $\pi^\pm$ and $\pi^0$ mass difference taking into account gauge coupling and casimir factors. Because this other sector is not identical to QCD, this estimate only gives a rough estimate for the mass of $\pi_A$. Lattice calculation can improve the precision of the estimate of the mass. The pion $\pi_A$ has couplings shown in Eq.~\ref{eq: octetdual-coupling}. It can be both singly produced and pair produced at the LHC, leading respectively to dijet resonances in dijet and four jet final states~\cite{Aad:2014aqa,ATLAS:2015nsi,Khachatryan:2015dcf,Khachatryan:2015sja,Khachatryan:2014lpa}.  

In Figure~\ref{fig:axion-pion} we show the current limits on the axion production cross section times branching ratio into photos~\cite{ATLAS:new,CMS:new} as a function of the axion mass for the example case $(Y_\psi, \, Y_\chi) = (0, \, 1)$~\footnote{The requirement that $\Lambda_a \geq \Lambda_{\rm QCD'}$ implies the relationship $m_a \leq 2.2 m_{\pi_A}$. This constraint is not shown in the figure since it is a region that is already ruled out by experiment.}. We also plot contours corresponding to fixed values of the scalar octet mass and contours of the production cross-section of the scalar octet as a function of the axion mass and production cross-section. Also shown are the exclusion limits on the scalar octet from dijet searches with 8 TeV LHC data~\cite{Aad:2014aqa,Khachatryan:2015sja}. One sees from Figure~\ref{fig:axion-pion} that clearly our model would be first discovered by detecting the axion in the diphoton channel. 

In Table~\ref{table:production decay} we present the production cross sections times the relevant branching ratios for both the axion and scalar octet in two benchmark points for our model. These benchmark points were chosen to reproduce the potential signal observed by ATLAS and CMS~\cite{ATLAS,CMS:2016owr} with 2015 data\footnote{
Following the announcement of this potential signal a large number of papers were written proposing models to fit the data, see e.g.~\cite{Franceschini:2016gxv,Strumia:2016wys} for a summary and a fairly complete list of the proposed models. The signal was not observed in their analysis of 2016 dataset (which had almost four times more data).
}, which is well fit by a $750$ GeV resonance with production times branching ratio to diphotons of about $6$ fb. Both of these benchmark points have recently been ruled out by the updated analysis~\cite{ATLAS:new,CMS:new}, but are still useful to illustrate general features of our model. We see from Eq.~\ref{eq: axion-partial-width} that the only difference in the axion phenomenology with a non-zero $Y_\psi$ is that it decreases the branching ratio into photons and $Z$s. Therefore non-zero $Y_\psi$ requires a larger overall axion production cross section to reproduce the same signal in the diphoton channel. This leads to a smaller mass for the scalar octet and increases its production cross section as can be seen in Table~\ref{table:production decay}, in addition to a non-zero branching ratio of $\pi_A$ to photon and one jet.

\begin{table}[t]
\centering
\begin{tabular}{c|c|c|c|c|c|c|} 
\multirow{2}{*}{Particle} & \multirow{2}{*}{$(Y_{\psi},Y_{\chi})$} & \multicolumn{5}{|c|}{$\sigma \times \text{BR}$} \\ \cline{3-7}
& & $\gamma\gamma$ & $gg$ & $\gamma Z$ & $ZZ$ & $g \gamma$ \tabularnewline
\hline 
\multirow{2}{*}{axion (13 TeV)} & (0,1) & $6 \,\text{fb}$  & $ 0.15 \,\text{pb} $  & $3.6 \,\text{fb}$ & $0.6 \,\text{fb}$ & - \\
  & (1/3,1) & $6 \,\text{fb} $ & $ 0.19 \,\text{pb} $ & $3.6 \,\text{fb}$ & $0.6 \,\text{fb}$ & - \\
\hline
\multirow{2}{*}{$\pi_A$ (8 TeV)}  & (0,1) & - & $0.2\,\text{pb}$  & - & - & - \\
 & (1/3,1) & - & $0.3\,\text{pb}$   & - & - & $13 \,\text{fb}$ \\
\hline
\multirow{2}{*}{$\pi_A$ (13 TeV)}  & (0,1) & - & $1.3\,\text{pb}$  & - & - & - \\
 & (1/3,1) & - & $1.8\,\text{pb}$   & - & - & $77 \,\text{fb}$ \\
\end{tabular}\caption{The single production times branching ratio of the two lightest pseudo-goldstones in our model. The axion production cross section is for $\sqrt{s} = 13$ TeV and the one for the color-octet scalar, $\pi_A$, is both for $\sqrt{s} = 8$ TeV and $\sqrt{s} = 13$ TeV.  We set $f_a = 480 (Y_\chi^2 - Y_\psi^2)$ GeV to match the diphoton anomaly and use the benchmark choices of hypercharge $(Y_{\psi},Y_{\chi}) = (0,1)$ and $(Y_{\psi},Y_{\chi}) = (1/3,1)$. The mass of the axion is 750 GeV while we took the mass of $\pi_A$ to be 1.1 TeV and 1 TeV for $(Y_{\psi},Y_{\chi}) = (0,1)$ and $(Y_{\psi},Y_{\chi}) = (1/3,1)$ respectively.}\label{table:production decay}
\end{table}

Pair produced color octets are searched for by CMS with 20 $\text{fb}^{-1}$ of 8 TeV data~\cite{Khachatryan:2014lpa}. The pair production cross-section is dominated by the usual coupling to gluons through the covariant derivative and thus depends only on the octet's mass. We recast this search to exclude the color octet $\pi_A$ with masses below $\sim 700 \,\text{GeV}$~\cite{GoncalvesNetto:2012nt} assuming it decays predominantly to dijets. Improved searches for pair produced color octet scalars with $13 \,\text{TeV}$ data  (see e.g.~\cite{Degrande:2014sta} for the scalar octet pair production cross section at the 13 TeV LHC) can complement the searches for single production in dijet events shown in Figure~\ref{fig:axion-pion}.

The color octet scalar can also decay into a photon and a jet if $Y_\psi \neq 0$.  The branching ratio into jet and photon can be approximated by
\beq
BR(\pi_A\rightarrow g \gamma) \approx  \left( \frac{ 24 e^2 Y_{\psi}^2}{24 e^2 Y_{\psi}^2 + 5 g_3^2 } \right) \approx 4.6 \%   \left( \frac{Y_{\psi}}{1/3} \right)^2 .
\eeq
If the axion is identified with the 750 GeV signal, the strongest constraint on $\sigma_{\pi_A} \times BR(\pi_A\rightarrow g \gamma)$ is at about $10 (13)\,\text{fb}$ with $20 \,\text{fb}^{-1}$ of 8 TeV LHC data at ATLAS and CMS~\cite{Aad:2013cva,Khachatryan:2014aka} at 1.1 TeV (1 TeV). This places a bound on a 1 TeV mass $\pi_A$ that is roughly
\beq
\left| Y_{\psi} \right| \leq 1/3.
\eeq
Our second benchmark case in Table~\ref{table:production decay} was chosen to saturate this constraint. This constraint is stronger than the one from the dijet resonance search for the benchmark case $Y_{\psi} = 1/3$ and $Y_{\chi} = 1$, and offers a signature that distinguishes various choices of hypercharge $Y_{\psi}$ and $Y_{\chi}$ in our model. This example illustrates the complementarity of the diphoton searches for the axion and the dijet and gamma jet searches for the octet in order to investigate our model at the LHC. There are new searches at 13 TeV from the ATLAS collaboration~\cite{Aad:2015ywd} for jet plus photon resonances which probe scalar octets decaying to photon and a jet with masses larger than $1.5$ TeV, and should probe interesting parameter space with more data.

\subsection{Other Heavy States}

The production of the heavy states in our model all require energies comparable or larger than the $SU(3)'_c$ confinement scale, and are therefore beyond the reach of the forthcoming LHC run. In this section, we discuss briefly the collider signatures of each type of heavy states in our model, which might be of interest in future colliders.

The rest of the pseudo-goldstone bosons ($\pi'_A$, $\Pi_A$, $\phi_A$ and $\phi'_A$) in our model are charged under the $SU(3)'_c$ and receive a loop generated mass of order $g'_3 f_a$. The $SU(3)'_c$ is already strongly coupled at $\text{TeV}$ energies and therefore the one-loop calculation of the mass of these pseudo-goldstones only provides a very rough estimate of their physical masses (shown in Table \ref{table:masses of states}). These pseudo-goldstones all carry $SU(3)'_c$ quantum number and therefore can only be pair produced.  Given the proximity of their masses to $\Lambda_{\rm QCD'}$, it may be more relevant to consider the mass of their bound states. Due to their large mass, these pseudo-goldstones will be produced near threshold at a collider. In this case they cannot develop a dark shower since there are no light $SU(3)'$ states for them to radiate. Instead they will be produced as a meson anti-meson bound state, analogous to charmonium states, and promptly annihilate to final states with many jets.

\begin{table}[t]
\centering
\begin{tabular}{c|c|c|c|c|c|c|c|c} 
& $a$ & $\sigma \times BR$ & $\pi_A$ & $\pi'_A$ & $\Pi_A$ & $\phi_A$ & $\overline{\phi}_A$ & $\eta'_A$ \tabularnewline
\hline 
Reps & $(1,1)$ & - & $(8,1)$ & $(1,8)$ & $(8,8)$ & $(3,\bar{3})$ & $(\bar{3},3)$ & (1,1) \tabularnewline
\hline
BM1 & 750 GeV & 6 fb & 1100 GeV & {\it 4} TeV & {\it 4} TeV & {\it 2} TeV & {\it 2} TeV & {\it 5} TeV
\tabularnewline
\hline
BM2 & 750 GeV & 0.2 fb & 5900 GeV & {\it 21} TeV & {\it 21} TeV & {\it 11} TeV & {\it 11} TeV & {\it 18} TeV
\tabularnewline
\hline
BM3 & 1400 GeV & 0.2 fb & 2000 GeV & {\it 8} TeV & {\it 8} TeV & {\it 3.5} TeV & {\it 3.5} TeV & {\it 6} TeV
\end{tabular}\caption{The representations and masses of the pseudo-goldstones in different benchmark models with $(Y_{\psi},Y_{\chi}) = (0,1)$.  The pseudo-goldstone masses were calculated at 1 loop, but the $SU(3)'_c$ charged pseudo-goldstone masses (in {\it italic} form) are very uncertain due to the strongly coupled $SU(3)'_c$ at the TeV scale. For the mesons charged under $SU(3)' _c$ with masses near or below the confinement scale of QCD$'$ ($\Lambda_{\rm QCD'} \approx (3.0,\, 6.7, \, 5.3) \,{\rm TeV}$ for the three benchmark points respectively), the more relevant quantities for collider physics are the masses and widths of their associated bound states.}\label{table:masses of states}
\end{table}

The $\eta'_A$ meson in our model, like the $\eta'$ meson in the Standard Model, receive a mass from non-perturbative effects of order the confinement scale of $SU(3)_a$. Its mass lies in the few TeV region for the range of parameters we are interested in. The $\eta'_A$ meson can be singly produced through the dimension five operator like the axion. It decays dominantly to two back to back jet and can be searched for in dijet resonances. The other mesons and baryons from the confinement of the $SU(3)_a$ have masses comparable to the $\eta'_A$ meson and lead to collider signatures similar to that of the pseudo-goldstone bosons.

There are also glueball states of both the confining group $SU(3)'_c$ and $SU(3)_a$. The glueball states have masses roughly 7 times the confinement scale of the two confining groups~\cite{Craig:2015pha}, at tens or hundreds of TeV.

\subsection{Cosmology}

The reheating temperature in our model cannot be above the confinement scale of QCD$'$ or $SU(3)_a$. There are several reasons for this. Below the $SU(3)_a$ confinement scale, the fermions that were charged under $SU(3)_a$ are removed from the spectrum and we are left with the mesons $\Pi_A, \, \pi'_A, \, \phi_A$ and $\overline{\phi}_A$ that are charged under $SU(3)'_c$. After $SU(3)'_c$ confines, $\phi_A$ and $\overline{\phi}_A$ can form baryon-like bound states $\phi_A^3$ and $\overline{\phi}_A^{3}$, with masses larger than $O(10 \,{\rm TeV})$. These bound states are stable because they carry non-zero $\psi$ and $\chi$ number, which are unbroken $U(1)$ symmetries. They also have electric charge $\pm 3 (Y_\psi - Y_\chi)$, which is necessarily non-zero for the axion to decay into photons (see Eq.~\ref{eq: axion-partial-width}).  Stable charged particles are heavily constrained by various Charged Massive Particles (CHAMP) searches.

Searches based on spectroscopy of heavy-particle concentration enriched water samples constrain their number abundance to be~(see e.g. \cite{Hemmick:1989ns,Verkerk:1991jf,Burdin:2014xma} and references therein)
\bea
n_{\rm CHAMP} \lesssim (10^{-20}  -  10^{-15}) n_p,
\eea  
where $n_{\rm CHAMP}$ and $n_p$ are number density of CHAMP and proton respectively. This number densities are orders of magnitude smaller than the expected thermal abundance of the baryon-like states $\phi_A^3$ and $\overline{\phi}_A^{3}$. These stable element bounds require that the reheating temperature is less than about a factor of 50 compared to the masses of the states $\phi_A^3$.

There is also the standard axion domain wall problem~\cite{Sikivie:1982qv}.  A quick analysis of our model shows that there are nine degenerate vacua in our theory due to the spontaneous symmetry breaking of discrete non-anomalous rotations of $\psi$ and $\chi$.  These stable domain walls would overclose our universe and are observationally excluded.  The reheating temperature needs to be lower than the confinement scale of QCD$'$.  This is a weaker constraint than the constraint from CHAMP searches.  In summary, the reheating temperature in our model has to be less than a few hundred GeV. 

\section{Conclusion}

We conclude by first discussing some previous approaches that are similar to what was considered in this paper.  The first heavy axion models were given in Ref.~\cite{Dimopoulos:1979qi, Dimopoulos:1979pp}.  These models were all technicolor models and have since been excluded.  The fact that $\mathbb{Z}_2$ solutions to the strong CP problem could give a visible axion around the TeV scale was first mentioned in Ref.~\cite{Hook:2014cda}.  In that model, the scalar color octet is lighter than the composite axion leading to a model with very different phenomenology.  Other $\mathbb{Z}_2$ models, e.g. Ref.~\cite{Rubakov:1997vp}, if used to make the axion collider visible, lead to fine tuning problems worse than the strong CP problem.  Solving these problems leads naturally the model presented in the paper.  

More recently, the fact that the QCD axion could be collider visible was proposed in Ref.~\cite{Gherghetta:2016fhp}.  By unifying QCD into a larger gauge group, they get an additional confining gauge group whose $\theta$ angle is the same as the QCD $\theta$ angle.  However, in all of their models they need to explicitly set a CP violating phase to zero to keep both $\theta$ angles equal.  Another recent proposal was to make the the radial mode of the field that breaks the PQ symmetry~\cite{Chiang:2016eav} visible at the LHC.  Both of these proposals introduce a new hierarchy problem by introducing a fundamental axion with $f_a \sim$ TeV.

We next discuss the issue of whether the $\mathbb{Z}_2$ symmetry sets the theta angles of the two sectors equal, see e.g. Ref.~\cite{Dvali:2005an} for concerned discussion.  In the Lagrangian formalism, the $\theta$ angles are equal due to the appearance of $\theta$ in the Lagrangian.  In the Hamiltonian formalism, $\theta$ appears instead as a parameter labeling different superselection sectors.  In order to see that the $\mathbb{Z}_2$ symmetry still requires that the theta angles are the same, it is sufficient to study the limit where the soft $\mathbb{Z}_2$ symmetry breaking has been set to zero.  In this limit the statement that the IR physics has a $\mathbb{Z}_2$ symmetry is simply the statement that the states and spectrum are $\mathbb{Z}_2$ symmetric.  As shown by Ref.~\cite{Vafa:1984xg}, two sectors with differing $\theta$ have different spectrum, e.g. 
the vacuum energies are different.  Thus the $\mathbb{Z}_2$ symmetry also implies that the $\theta$ of the two sectors must be the same.  Turning on soft symmetry breaking masses for the Higgses does not change the arguments.

In this article, we proposed that the QCD axion itself could be visible at the LHC.  By using conformal dynamics and strong CP considerations, we have found a model where all of the new physics is expected to appear at the TeV scale.  In order to solve the strong CP problem, massless quarks are needed, which renders this theory extremely predictive.  Lattice calculations are important to determine the exact mass of the scalar color octet.

\acknowledgments
We thank Tony Gherghetta, Kiel Howe, Jeremy Mardon, Leonard Susskind and Ken Van Tilburg for very valuable discussions. We thank NSF grant PHYS-1316699 for support. A.H. and G.M.T. are also supported by the DOE Grant DE-SC0012012. A.H and G.M.T acknowledge the Aspen Center for Physics, which is supported by National Science Foundation grant PHY-1066293, where part of this work was completed. We thank in advance Gia Dvali for spirited future discussions on Ref.~\cite{Dvali:2005an}.

\appendix

\bibliography{reference}
\bibliographystyle{JHEP}

\end{document}